\documentclass[final]{svjour3}

\usepackage[fleqn]{amsmath}
\usepackage{amsmath}
\usepackage{eucal}
\usepackage{amssymb}
\usepackage{mathrsfs}
\usepackage{graphicx}
\usepackage{cite}
\usepackage{authblk}
\usepackage{multirow}
\usepackage{subfig}

\usepackage[normalem]{ulem}

\begin{document}

\title{\Large{\textbf{Computational Modeling of an MRI Guided Drug Delivery System Based on Magnetic Nanoparticle Aggregations  for  the Navigation of Paramagnetic  Nanocapsules }}}

\author{N.K. Lampropoulos\thanks{Corresponding author: N.K. Lampropoulos, Email : nikolaoslampropoulos@hotmail.com, tel : +30 697 93 86 453, fax : +30 210 53 85 306 }  \and E.G. Karvelas \and I.E. Sarris}

\institute{Department of Energy Technology, Technological \& Educational Institute of Athens, Agiou Spiridonos 17, 12210 Athens}

\titlerunning{Computational Modeling of an MRI Guided Drug Delivery System.}

\maketitle
\begin{abstract}
 A computational method for magnetically guided drug delivery is presented and the results are compared for the aggregation process of magnetic particles within a fluid environment. The  model is developed for the simulation of the aggregation patterns of magnetic nanoparticles  under the influence of MRI magnetic coils. A novel approach for the calculation of the drag coefficient of aggregates is presented. The comparison against experimental and numerical results from the literature is showed that the proposed method predicts  well the aggregations in respect to their size and pattern dependance, on the concentration and the strength of the magnetic field, as well as their velocity when particles are driven through the fluid by magnetic gradients.
\end{abstract}

\section{INTRODUCTION}
From the beginnings of 1970s researchers were studying the concept of magnetic guided drug delivery method~\cite{first,second}. The concept of this method is to attach the drug to the micro- or nanoparticles and then to inject them to the bloodstream. For the guidance to the targeted area, a Magnetic Resonance Imaging (MRI) device is needed. By making use of the magnetic guided drug delivery method, the quantity of the drug
required to reach therapeutic levels is being reduced. Also, the drug concentration at targeted sites is increased.

 The efficiency of this concept depends on the materials of the particles (cobalt and manganese ferrites, encapsulated or coated~\cite{Llan10}), the blood flow rate and the intensity of the magnetic field~\cite{third}.  It is found that, this method is more efficient in small blood vessels with low blood flow rates than large blood vessels, where the flow rate is higher. A small size of particles implies a magnetic response of reduced strength and as a result, it is difficult to drive particles and keep them in the targeted region~\cite{fourth}.
In order to overcome this difficulty the use of magnetic particles to create aggregations is proposed in order to increase the magnetic response~\cite{sixth}. The total magnetic moment of clusters is higher than that of isolated particles, and therefore, cluster are more magnetically responsive~\cite{ju1}. The ratio of the velocity of the aggregates to the velocity
of an individual particle was
found to reach a constant value independent of the aspect ratio
value~\cite{net08}.
 The magnetization  of nanospheres increases as a result of the increase in the concentration of matter along the direction of the field. This increase leads
to the observed change in velocity~\cite{net13}.
When the aggregations reach the targeted region, they break up into single particles. This break up can be achieved by using superparamagnetic particles that lose the magnetization after moving out from the magnetic field~\cite{sixth}. The size of aggregates is a very important parameter, since large aggregations could form clots in small arteries. On the other hand, a small aggregation would be dragged away from the cardiovascular system circulation. Therefore, the size of aggregation has an important role in the effort to create an efficient propulsion through the blood vessels. 
Magnetic drug targeting is possible in treating superficially located diseases, such as tumors and vascular diseases.  The treatment of tissues located away from the body surface is not possible because the magnetic force decrease with the increasing distance from the electromagnetic coil~\cite{Gl07}. In order to overcome this difficulty magnets are implemented near the targeted sites~\cite{seventh,eighth}.

Since an analytical study of aggregation process is impossible to be developed, a numerical model for magnetically guided drug delivery is attended. 
The method that is proposed here simulates the movement of aggregated magnetic particles in a fluid environment. The numerical model can simulate the number of resulting aggregations whose size and pattern depend on the concentration  and the strength of the magnetic field. The forces acting on a particle are described in detail in Section $2$. The model is validated through a comparison with two benchmark test cases of Ref.~\cite{sixth} and Ref.~\cite{nineth} and the results are discussed in Section $3$. Finally, discussion and conclusions are presented in Sections $4$ and $5$, respectively.

\section{METHODS}
For the propulsion model of the particles, six major forces are considered, i.e.
the magnetic force from MRI’s Main Magnet static field as well as the Magnetic field gradient force from the special Propulsion Gradient Coils. The static field caters for the aggregation of nanoparticles while the magnetic gradient navigates the agglomerations.
Moreover, the contact forces among the aggregated nanoparticles and the wall, and the Stokes drag force for each particle are considered, while only spherical particles are used here.
Finally, gravitational forces due to gravity and the force due to buoyancy are added.

 The motion of particles are given by the Newton equations:  \begin{equation}
 m_i\frac{\partial\boldsymbol{u}_i}{\partial t}=\boldsymbol{F}_{mag\_i}+\boldsymbol{F}_{nc\_i}+\boldsymbol{F}_{tc\_i}+\boldsymbol{F}_{drag\_i}+\boldsymbol{F}_{boy\_i}+\boldsymbol{W}_i 
\end{equation}
\begin{equation}
\boldsymbol{I}_i\frac{\partial \omega_i}{\partial t}=\boldsymbol{M}_{drag\_i}+\boldsymbol{M}_{con\_i}+\boldsymbol{T}_{mag\_i}
\end{equation}
where, the index $i$ stands for the particle $i$. The bold variables are vectors. The velocity is $\boldsymbol{u}_i$ and the rotational velocity is $\boldsymbol{\omega}_i$. The mass of particle $i$ is $m_i$, $t$ stands for the time and the mass moment of inertia matrix is $\boldsymbol{I}_i$. The linear and angular acceleration are $\frac{\partial\boldsymbol{u}_i}{\partial t}$ and $\frac{\partial \omega_i}{\partial t}$, respectively. The total magnetic force is $\boldsymbol{F}_{mag\_i}$. $\boldsymbol{F}_{nc\_i}$ and $\boldsymbol{F}_{tc\_i}$ are the normal and tangential contact forces, respectively. $\boldsymbol{F}_{drag\_i}$ stands for the hydrodynamic drag force. $\boldsymbol{F}_{boy\_i}$ and $\boldsymbol{W}_i$ are the buoyancy and the weight forces. $\boldsymbol{M}_{drag\_i}$ and $\boldsymbol{M}_{con\_i}$ stand for the drag and the contact moments, respectively. $\boldsymbol{T}_{mag\_i}$ is the torque due to the magnetic field at the position of particle $i$.

In the following section, the major forces that are taking into account during the simulation are described in detail.

        \subsection{\textit{Forces acting on particles}}

        \begin{enumerate}
        \item \textit{Magnetic Forces:}  Magnetic forces, $\boldsymbol{F}_{mag\_i}$, exerted on the $i$ particle are given by \cite{twelfth,thirteenth,fourteenth} $:$
\begin{equation}
\boldsymbol{F}_{mag\_i}=\boldsymbol{F}_{intmag\_i}+\boldsymbol{F}_{ismag\_i}
\end{equation}

where, $\boldsymbol{F}_{intmag\_i}$ is the magnetic force due to the interaction of the $i$th particle with the magnetic field, and $\boldsymbol{F}_{ismag\_i}$ is the magnetic force acting on the $i$th particle due to its interaction with the surrounding magnetic particles.

The force $\boldsymbol{F}_{intmag\_i}$ in the volume of the ferromagnetic body, $V,$ is given by: 
\begin{equation}
\boldsymbol{F}_{intmag\_i}=V(\boldsymbol{m}_i\cdot\nabla)\boldsymbol{B}_{ext\_i}
\end{equation}

The force $\boldsymbol{F}_{ismag\_i}$ is given by: 
  \begin{equation}
\boldsymbol{F}_{ismag\_i}=\sum\limits_j^N\boldsymbol{F}_{ismag\_ji}
\end{equation}

The numerical model for the magnetic forces is given in  \cite{nineth}.

\item\textit{Fluid Forces:} Each particle is subjected to the drag force that reads as:
 \begin{equation}
\boldsymbol{F}_{drag\_i}=\frac{1}{2}\rho u^2 C_d A
\end{equation}
where, $\rho$ is the density of the fluid, $u$ is the speed of the particle $i$ relative to the fluid and A is the reference area which equals to $\pi r^2$, where r stands for the radius of spherical particle. $C_d$ is the drag coefficient given by:
  \begin{equation}
\boldsymbol{C}_d=\frac{24 \left[1 + 0.15 Re^{0.687}\right]}{Re}
\end{equation} 
where, $Re$ is the Reynolds number based on the particle diameter \cite{fifteenth}.

In the present work, a new model is developed for the substitution of the reference area $A$, in Eq. $(6)$, with the effective area $A_{eff}$. This substitution takes into account the fact that the downstream spheres of the chain lie partially in the wake of the upstream one, therefore the reference area $A$ in Eq. $(6)$ of each sphere must be reduced to each real area that is exposed to the flow. On this basis  the area $A_{eff}$ can be calculated as follows $:$

\begin{figure}
\begin{center}
\includegraphics[bb=0 0 406 308,scale=0.35]{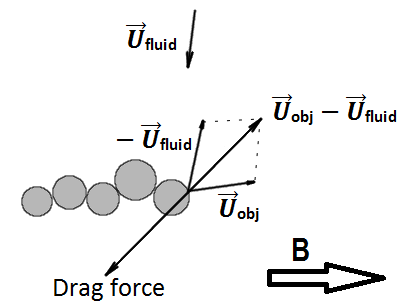}
\caption{Relevant velocity and drag force of a particle.}
\label{Drag force}
\end{center}
\end{figure}

\begin{enumerate}
\item Sweep all spheres.
\item Find the tangent sphere to the current one.
\item A straight chain assumption considered where the sphere $(1)$ is tangent only to two spheres namely the upwind and downwind. Sphere $(2)$ which lies upstream to sphere $(1)$ is found out by checking the inner product $ \overline{32} * (\vec{U}_{obj} - \vec{U}_{fluid}) $ to be positive, see Figs. \ref{Drag force} and \ref{Drag1}a. Where, $\vec{U}_{obj}$ stands for the velocity of the sphere and $\vec{U}_{fluid}$ stands for the velocity of the fluid 
\item Project the area of spheres (2 and 1) on the plane $\vec{U}_{obj} - \vec{U}_{fluid}$ that is cross  sphere ($2$) at each center, see Fig.~\ref{Drag1}b. This yields into two intersected circles, as shown in Fig. \ref{Drag3}. 
\item Change to plane coordinate system $(2D)$.
\item Calculate the overlapping area (white domain in Fig. \ref{Drag3}). The mathematical formula reads  as$:$ 
\begin{equation}
 E_{overlapping} =domain (234) + domain (134) - E(1234)
 \end{equation}
\item Substitute Eq. ($6$) with the equation: 
\begin{equation}
\boldsymbol{F}_{drag\_i}=\frac{1}{2}\rho u^2 C_d A_{eff}
\end{equation}
In this way, the real reference area $A_{eff}$ of sphere exposed to the flow, i.e.:
\begin{equation}
 A_{eff} = \pi {r_i}^2 - E_{overlapping} 
\end{equation} 
is taken into account.
\end{enumerate}
  \item\textit{Collision Forces:} Each particle interacts with other particles through contact.  In our model we use the Discrete Element Method (DEM) to calculate the motion of particles \cite{twentieth}.  DEM is a numerical model capable of describing the mechanical behaviour of assemblies of spheres and computing the motion and shear effect of a large number of small particles. In DEM, particles are approximated as rigid bodies and the interactions between them are explicitly considered \cite{seventeenth}.

\begin{figure}
\begin{center}
\subfloat[]{\includegraphics[width = 6.5cm]{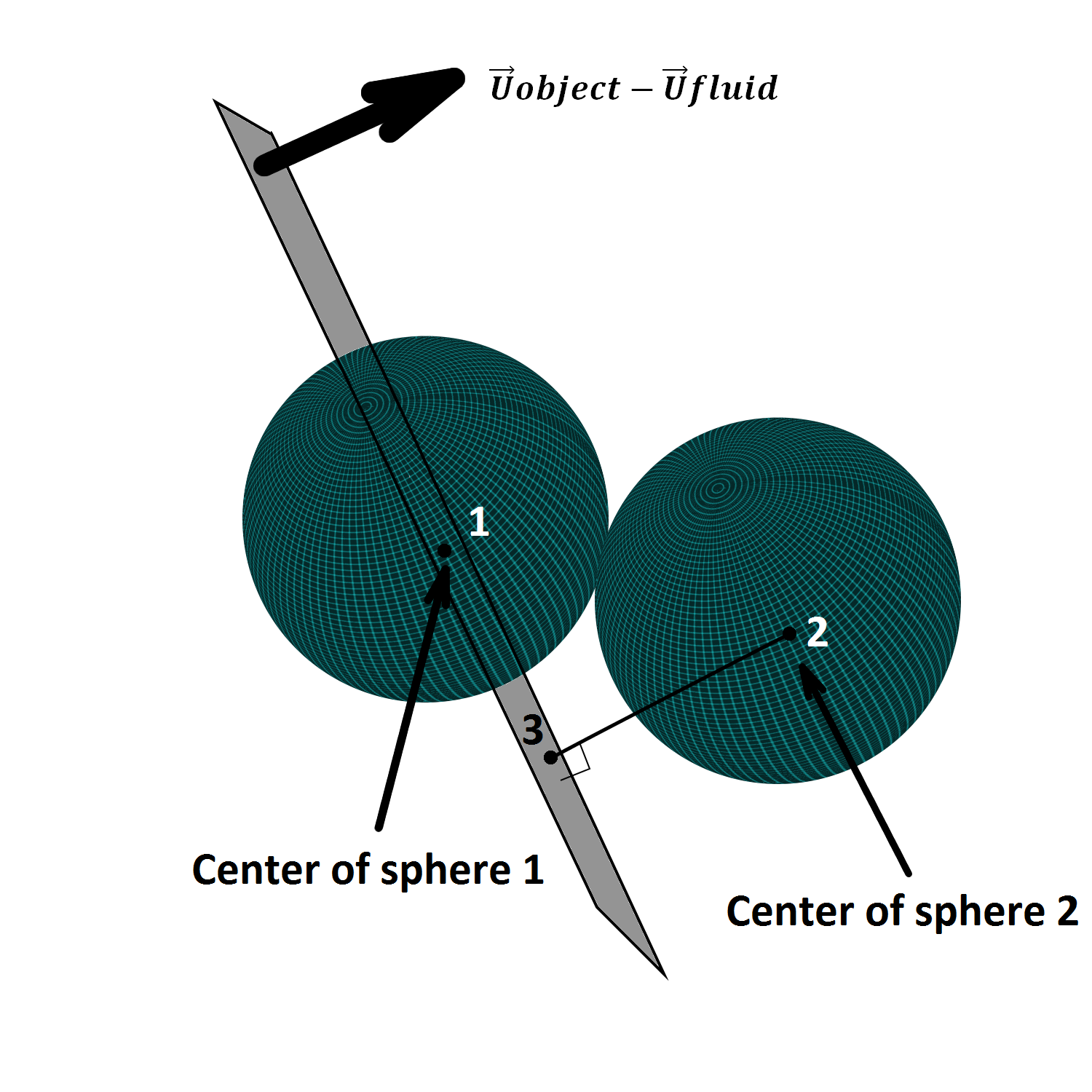}}
\subfloat[]{\includegraphics[width = 6.5cm]{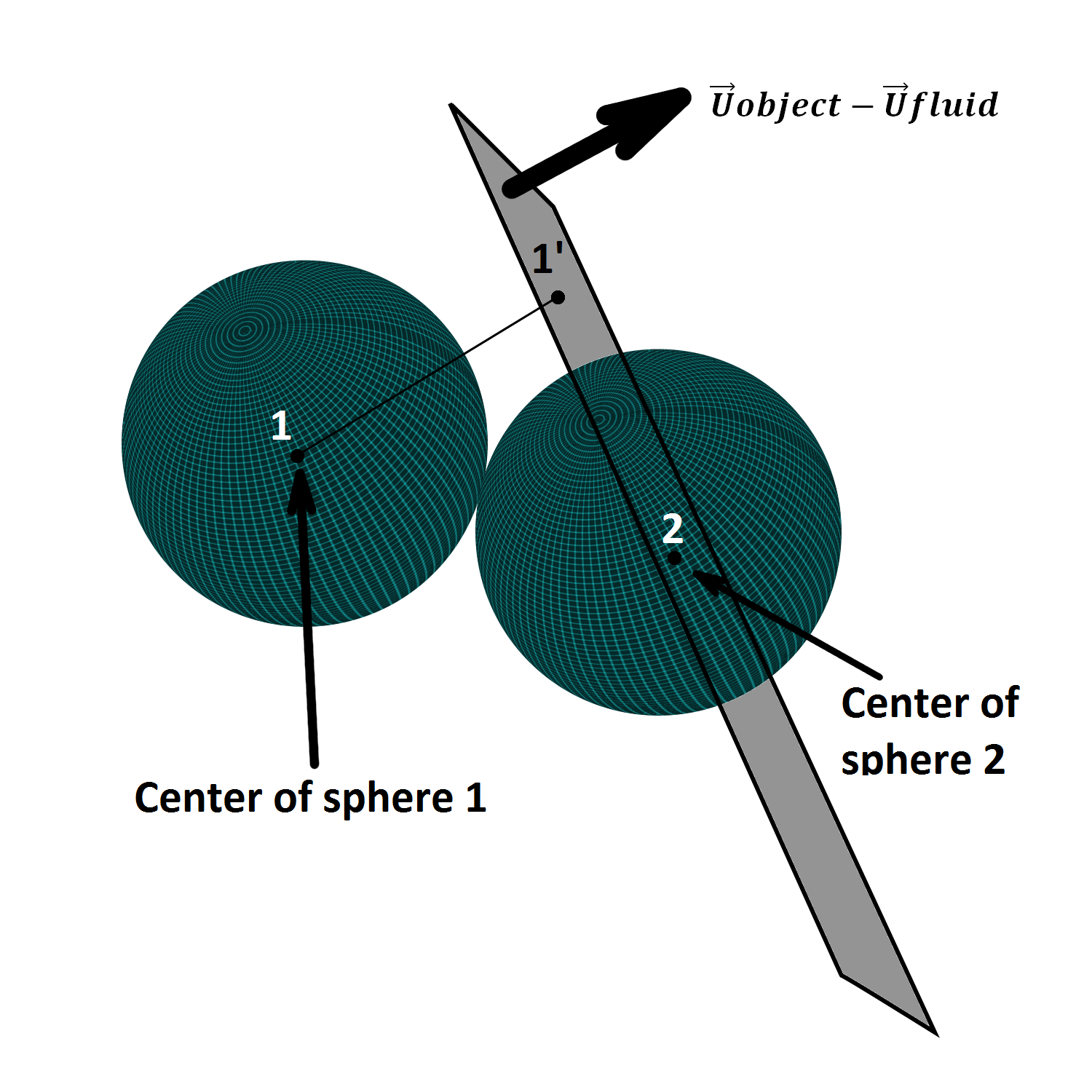}}
\caption{a) $3$ $\left(X_3,Y_3,Z_3\right)$ stands for the projection of sphere center $2$ $\left(X_2,Y_2,Z_2\right)$) on the plane $\vec{U}_{obj}-\vec{U}_{fluid}$ crossing sphere $1$ at each center $\left(X_1,Y_1,Z_1\right)$, and b) $1'$ $\left(X_{1'},Y_{1'},Z_{1'}\right)$ stands for the projection of sphere center $1$ $\left(X_1,Y_1,Z_1\right)$) on the plane $\vec{U}_{obj}-\vec{U}_{fluid}$ crossing sphere $2$ at each center $\left(X_2,Y_2,Z_2\right)$.}
\label{Drag1}
\end{center}
\end{figure}

\begin{figure}
\begin{center}
\includegraphics[bb=0 0 504 319,scale=.2]{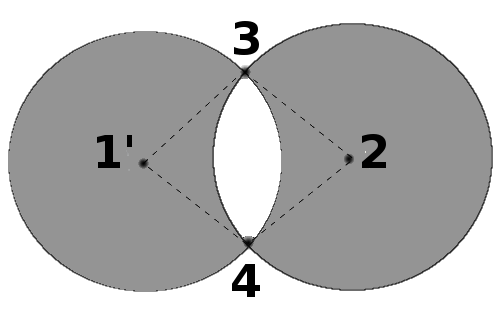}
\caption{Circular sector of two spheres  $E_{overlapping} =domain (234) + domain (134) - E(1234)$.}
\label{Drag3}
\end{center}
\end{figure}

\item\textit{Body Forces:} Both gravity and buoyancy are being included in the calculation of the body force. The force is addressed by  
\begin{equation}
\boldsymbol{F}_{grav\_i} = \boldsymbol{W}_i+\boldsymbol{F}_{boy\_i}=\frac{4}{3}\pi r_i^3(\rho_i - \rho_f)\boldsymbol{g}
\end{equation}
where, $\rho_i$ and $\rho_f$ are the density of the particle $i$ and the fluid, respectively, $r_i$ is the radius of the particle $i$ and $\boldsymbol{g}$ is the acceleration due to gravity.

\end{enumerate}  

\subsection{\textit{Magnetic field and interaction domain }}

The magnetic field $\boldsymbol{B}$ in the MRI bore is given by 
\begin{equation}
\boldsymbol{B}=\boldsymbol{B}_0 + \boldsymbol{\tilde{G}} +\boldsymbol{B}_1
\end{equation}
where, $\boldsymbol{B}_0$ is the MRI superconducting magnet field that is constant and uniform, $\boldsymbol{\tilde{G}}$ is the  gradient field and $\boldsymbol{B}_1$ is the time dependent radio frequency field \cite{sixteenth}.

  It is known that, the magnetic interaction force, $\boldsymbol{F}$, in the parallel direction between two spheres is inversely proportional to the fourth power of the separation distance \cite{tenth,eighteenth}:  \begin{equation}
\boldsymbol{F}\propto \frac{M_iM_j}{ (h + a_i + a_j)^4}
\end{equation}  
 where, $M_i$ and $M_j$ are the magnetic moments of the center of each sphere and $h$ is the nearest distance between the surfaces of two spheres with radii $a_i$ and $a_j$.
 
\begin{figure}
\begin{center}
\includegraphics[bb= 0 0 1844 1087, scale=.14]{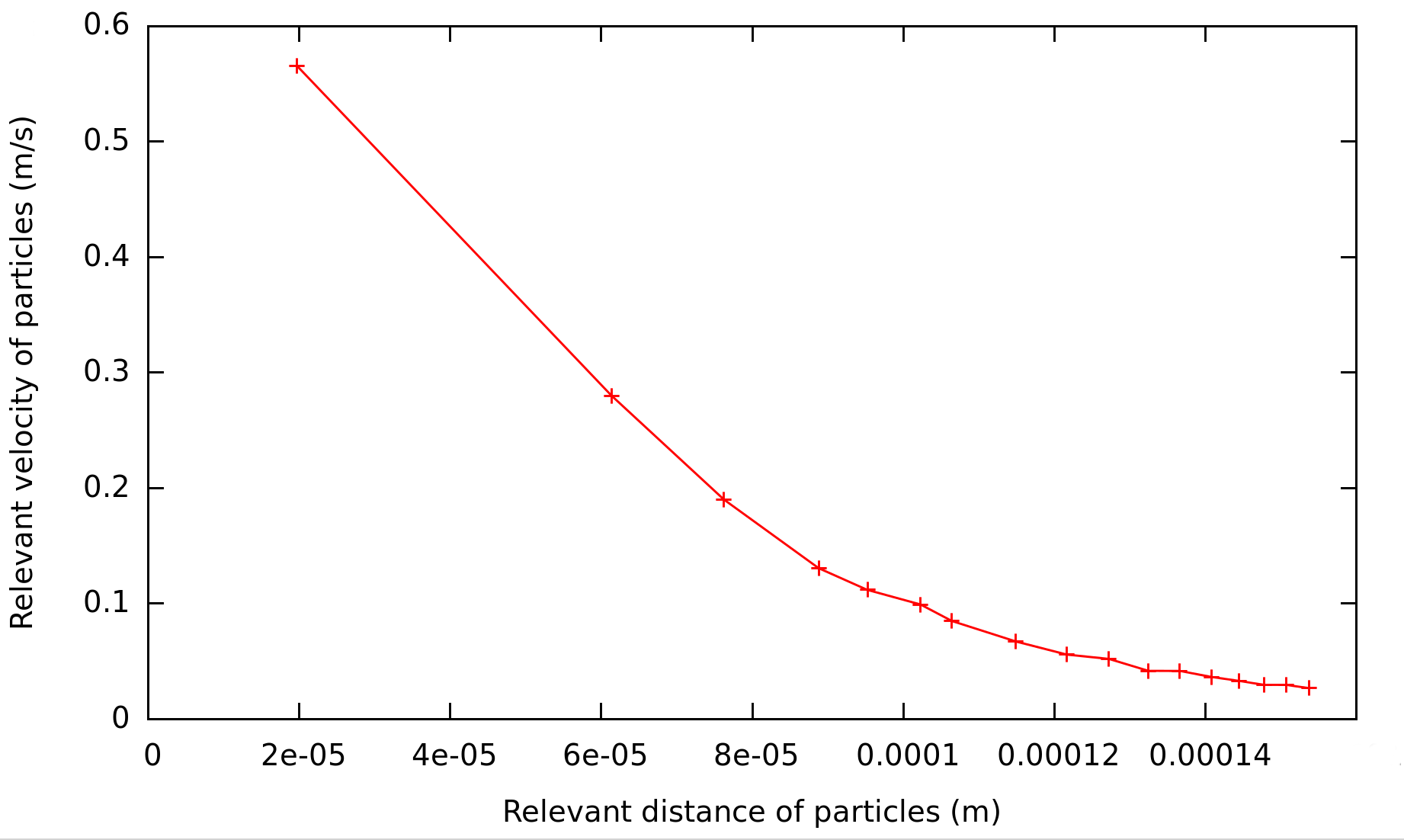}
\caption{Velocity of particles when approach under the force of a $1.5~T$ constant horizontal magnetic field from initial distance of 15 particle diameters of $11~\mu m.$ }
\label{15diameter}
\end{center}
\end{figure}

Although, a weak interaction is being observed for particles that are more than five radii apart,  as time goes by, the interaction force is getting stronger, because the particles are getting faster close to each other, as shown in Fig.~\ref{15diameter}, for the velocity between two approaching particles. Each particle interacts with all the other which are in the same area. The magnetic interaction domain is defined by the area of the domain that the particles are located in.
This is a time consuming method, because every time the magnetic moments of all particles are calculated. In order to accelerate the method, the magnetic moments that are exerted on particle $i$ are calculated from particles that are located in distance of 15 diameters. Then, only the magnetic moments that are located in distance of 10 diameters are stored. In this way, the magnetic moments of particles that are located in distance of 10 diameters are automatically recalled in order to be used for the calculation of the magnetic moments of other particles. Simulations of 64 and 125 particles of the method which was described above show agreement in half time with the method of calculating the magnetic moments exerted on particle $i$ from all the other particles.

\subsection{\textit{Numerical Method}}
 The OpenFoam platform was used for the calculation of the flow field and the uncoupled equations of particles motion \cite{twentieth}.
The simulation process reads as follow:
Initially, the fluid flow is found using the incompressible Navier-Stokes equations and the pressure correction method. Upon finding the flow field (pressure and velocity) the motion of particles is evaluated by the Lagrangian method by solving Eqs.~(1) and (2) along the trajectory of each particle.
The equations are solved in time by the Euler time marching method. The stability of the algorithm is guaranteed through a time step of the order of $10^{-6} s$.

The present numerical methods were validated against the results from Refs.~\cite{sixth,nineth}. In order to perform the comparisons, two series of simulations with the following computational domains and grid distributions were selected:  a) In the first case, four water solutions with different concentrations  (0.563, 1.125, 2.25 and 4.5 in $mg/ml$) and by using about $300-400$ particles  were simulated under the magnetization of a uniform magnetic field of $0.4~T$ in a stationary fluid. The spacing of the $3D$ computational grid of this  case was $2~ \times$ diameter of the particle in each direction. b) In the second case, a solution with concentration of 25 $mg/ml$ of $55$ polystyrene magnetic particles with diameter  of 5.5 $\mu m$ was simulated in distilled stationary water under a constant magnetic field of $0.005T$ and a gradient magnetic field of $\tilde G=1.4~ T/m$ along the $x-$axis. The spacing of the $2D$ domain  was $5~ \times$ diameter of the particle in each direction. The summary of the domain parameters for the first and the second case is shown in Table \ref{table6}.

\begin{table}
\caption{Simulation parameters}
\label{table6}
\centering
\begin{tabular}{|c|c|c|c|c|c|c|c|}
\hline
\multicolumn{8}{ |c| }{ Flow domain and computational grid of Case 1 (3D)} \\
\hline
No & Concentr. & Particles & Volume  & \multicolumn{3}{ |c| }{Dimensions in $(m)\times 10^{4}$  }  & Number of  \\
  &  ($mg/ml$)&  &  ($m^3$)  & \multicolumn{3}{ |c| }{ of $x\times y \times z$ directions}  &  cells (x,y,z) \\
\hline
1 & 0.563  & 311  & $4.185\times 10^{-10}$ & $7.48$ & $7.48$ & $7.48$ & $34 \times 34 \times  34$ \\
\hline
2 & 1.125  & 385  & $2.596\times 10^{-10}$ & $6.38$ & $6.38$ & $6.38$ & $29 \times 29 \times  29$ \\
\hline
3 & 2.25  & 336  & $1.133\times 10^{-10}$ & $4.84$ & $4.84$ & $4.84$ & $22 \times 22 \times  22$ \\
\hline
4 & 4.5  & 307  & $5.174\times 10^{-11}$ & $3.96$ & $3.96$ & $3.30$ & $18 \times 18 \times  15$ \\
\hline
\multicolumn{8}{ |c| }{ Flow domain and computational grid of Case 2 (2D)} \\
\hline
5 & 25  & 55  & $2.038\times 10^{-13}$ & $1.924$ & $1.924$ & $0.055$ & $7 \times 7$   \\
\hline
\end{tabular}
\end{table}

\section{RESULTS}
Two series of simulations were performed according to the parameters that are described in the previous section.  In the first case, which is initially addressed by Mathieu and Martel \cite{sixth}, the  aggregation of $Fe_3O_4$ particles dispersed in a stationary distilled water is investigated under the influence of a constant magnetic field.
  The density of the particle and the fluidic environment  was $1087$ $kg/m^3$ and $1000$ $kg/m^3$, respectively, while the diameter of the particles was considered to be 11 $\mu m$. The relative magnetic permeability of the $Fe_3O_4$ particles was $12.3$ and the magnetic permeability of the fluid was $4 \pi \times 10^{-7} ~A/m$. The Young's modulus of the material was $10^9~N/m^{2}$, the tangential stiffness was $10~Nsm^{-1}$ and the coefficient of friction and the Poisson ratio was $0.5$ for the two properties.

\begin{figure}
\begin{center} 
 \includegraphics[bb=0 0 1520 1074,scale=.15]{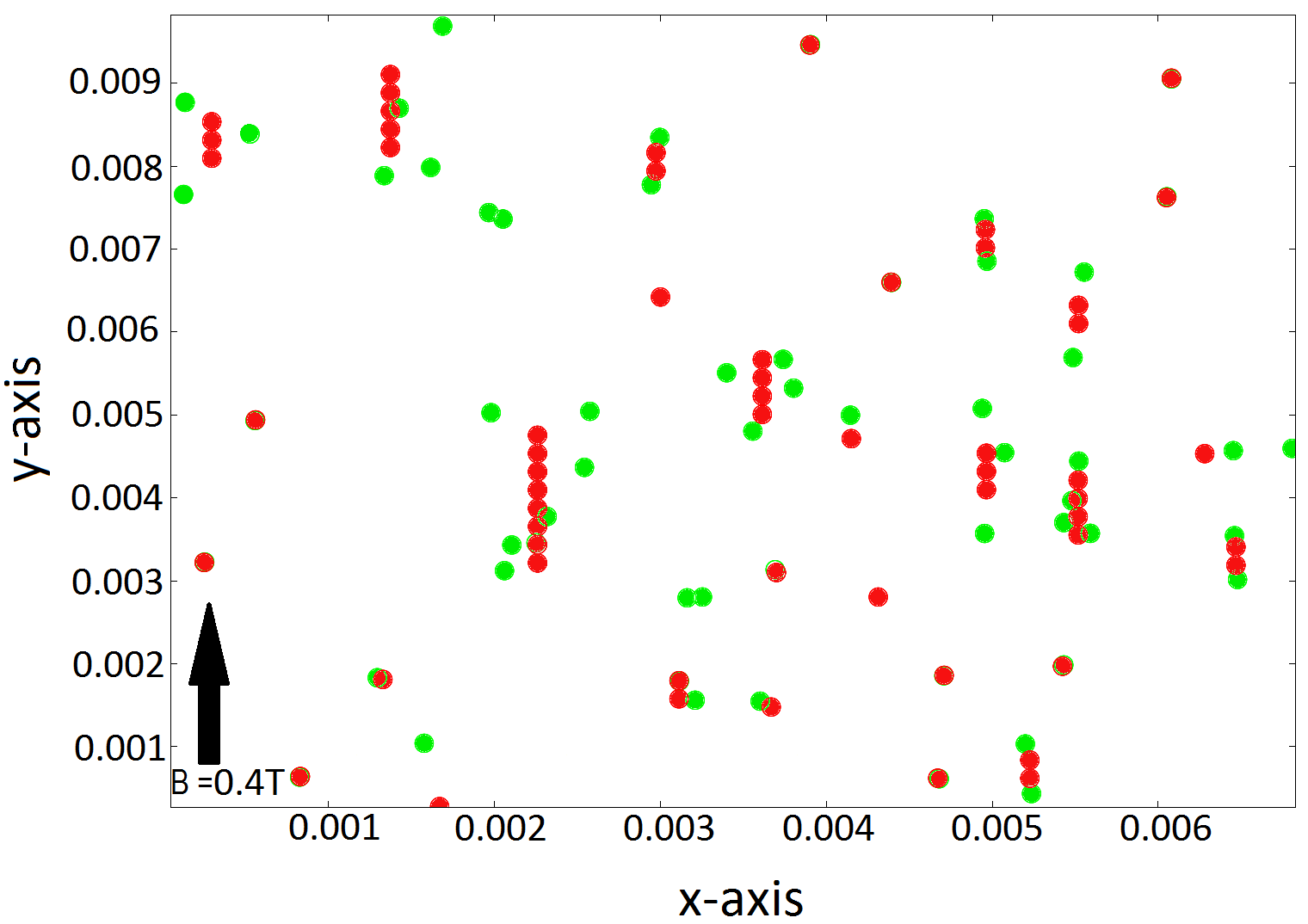}
 \caption{Green (lighter) particles are shown the initial particle position at $t=0s$, while red (darker) particles correspond to $t= 5~ms$ where the particles chains are fully formed parallel to the steady magnetic field.}
\label{chains}
\end{center}
\end{figure}

Four water solutions with particle concentrations of $0.563~mg/ml$, $1.125~mg/ml$, $2.25~mg/ml$ and $4.5~mg/ml$ were simulated under a uniform transverse magnetic field of magnitude $B_0=0.4~T$. No external magnetic gradient is applied during this simulation and the initial positions of the microparticles are randomly generated. At the beginning of each simulation, for $t=0s$, the magnetic field is $B_0~=~0~T$ and under the influence of the uniform magnetic field the particles (magnetic dipoles) are formed into chains that are oriented parallel to the  magnetic field within $t~=~5~ms$, as it is shown in Fig.~\ref{chains}.

\begin{figure}
\centering
 \subfloat[]{\includegraphics[width =6cm]{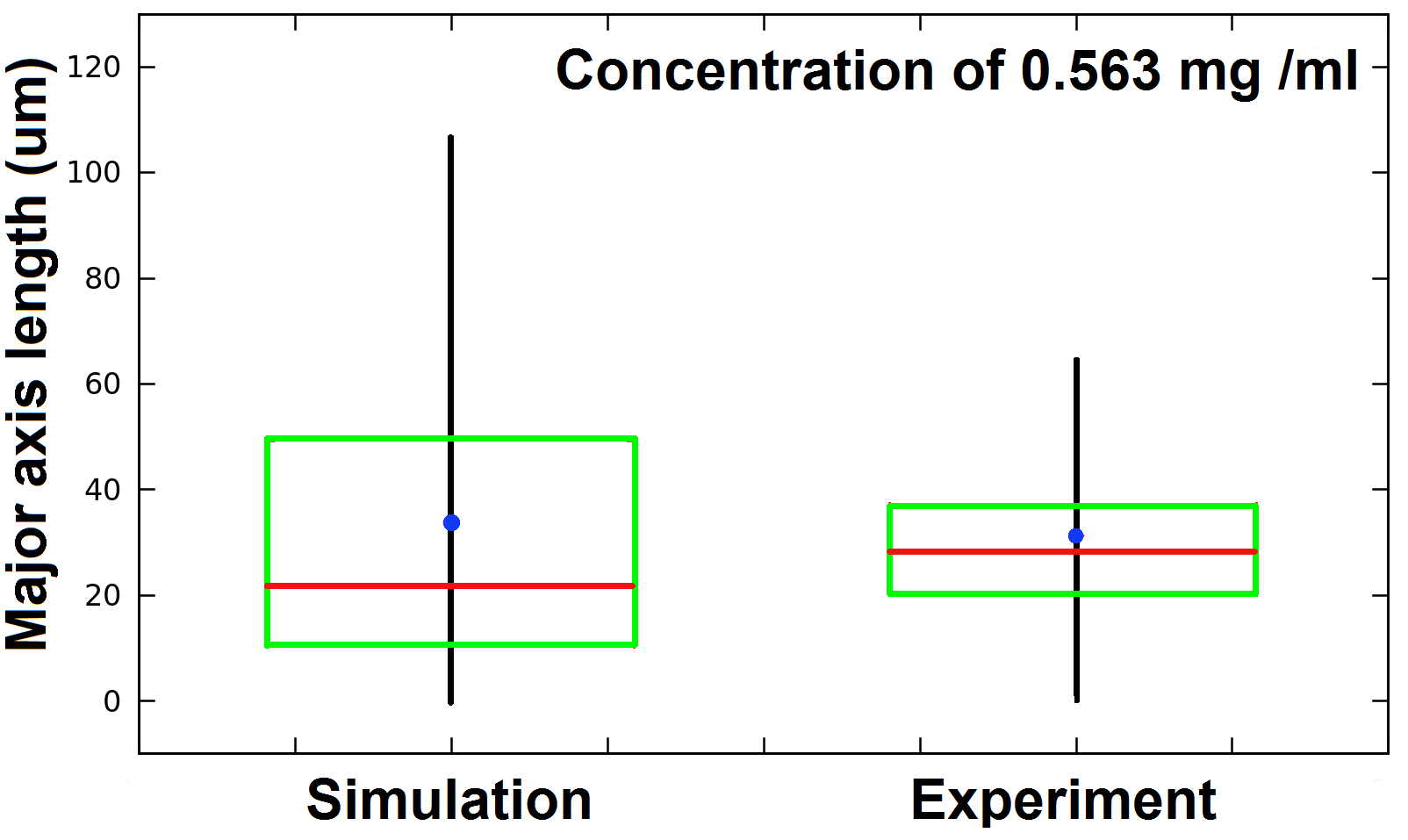}} 
 \subfloat[]{\includegraphics[width = 6cm]{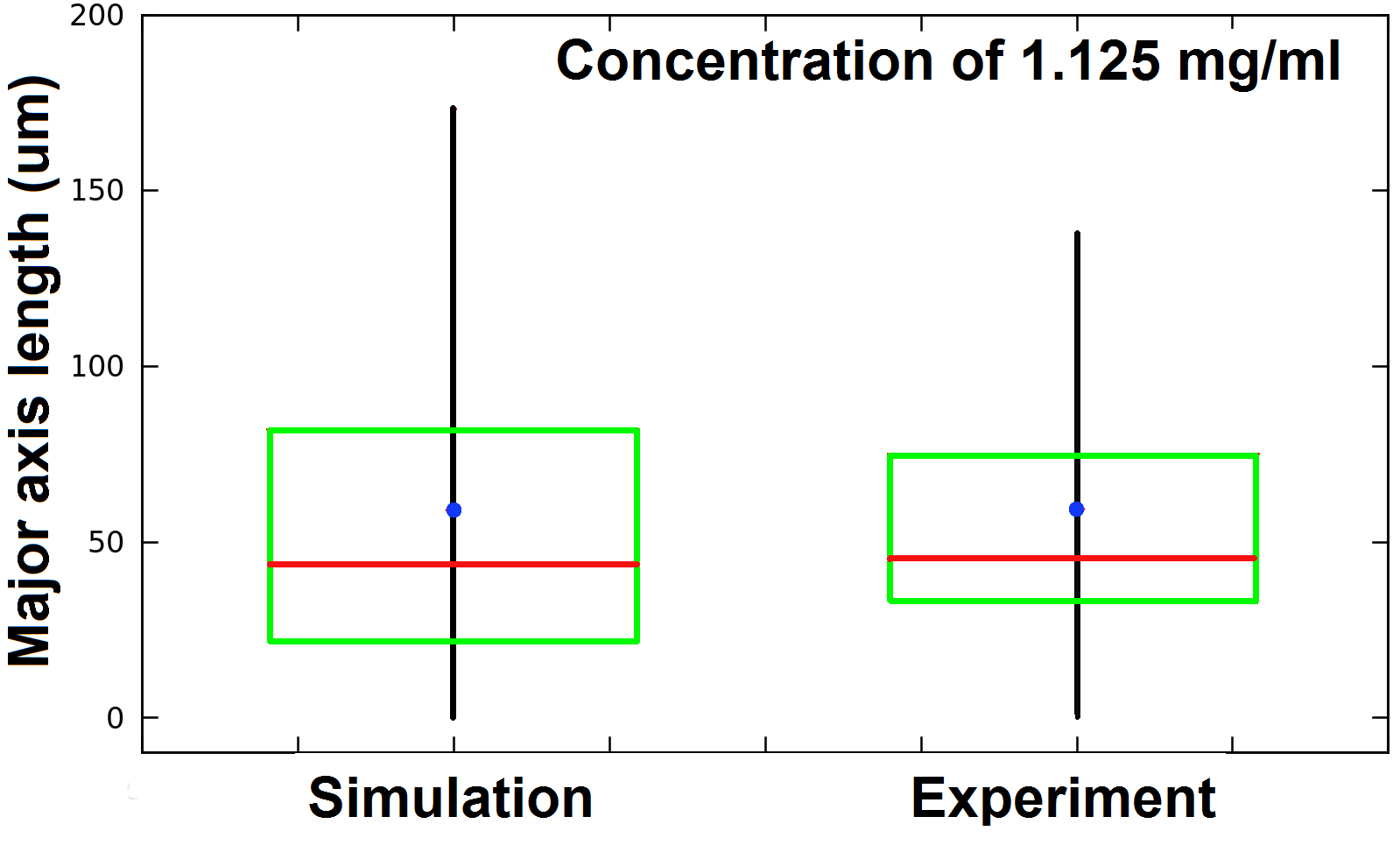}}\\
 \subfloat[]{\includegraphics[width = 6cm]{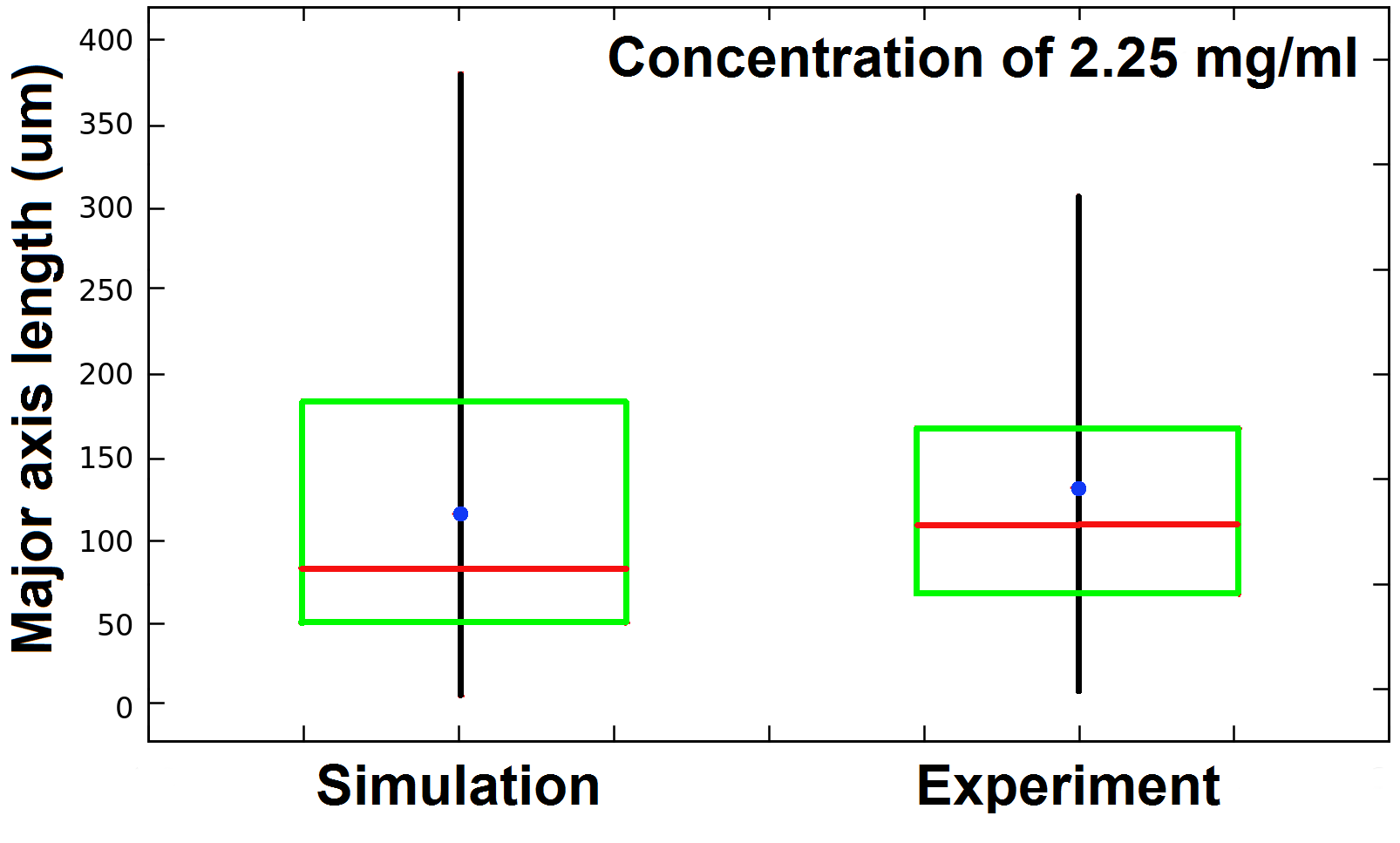}}
 \subfloat[]{\includegraphics[width = 6cm]{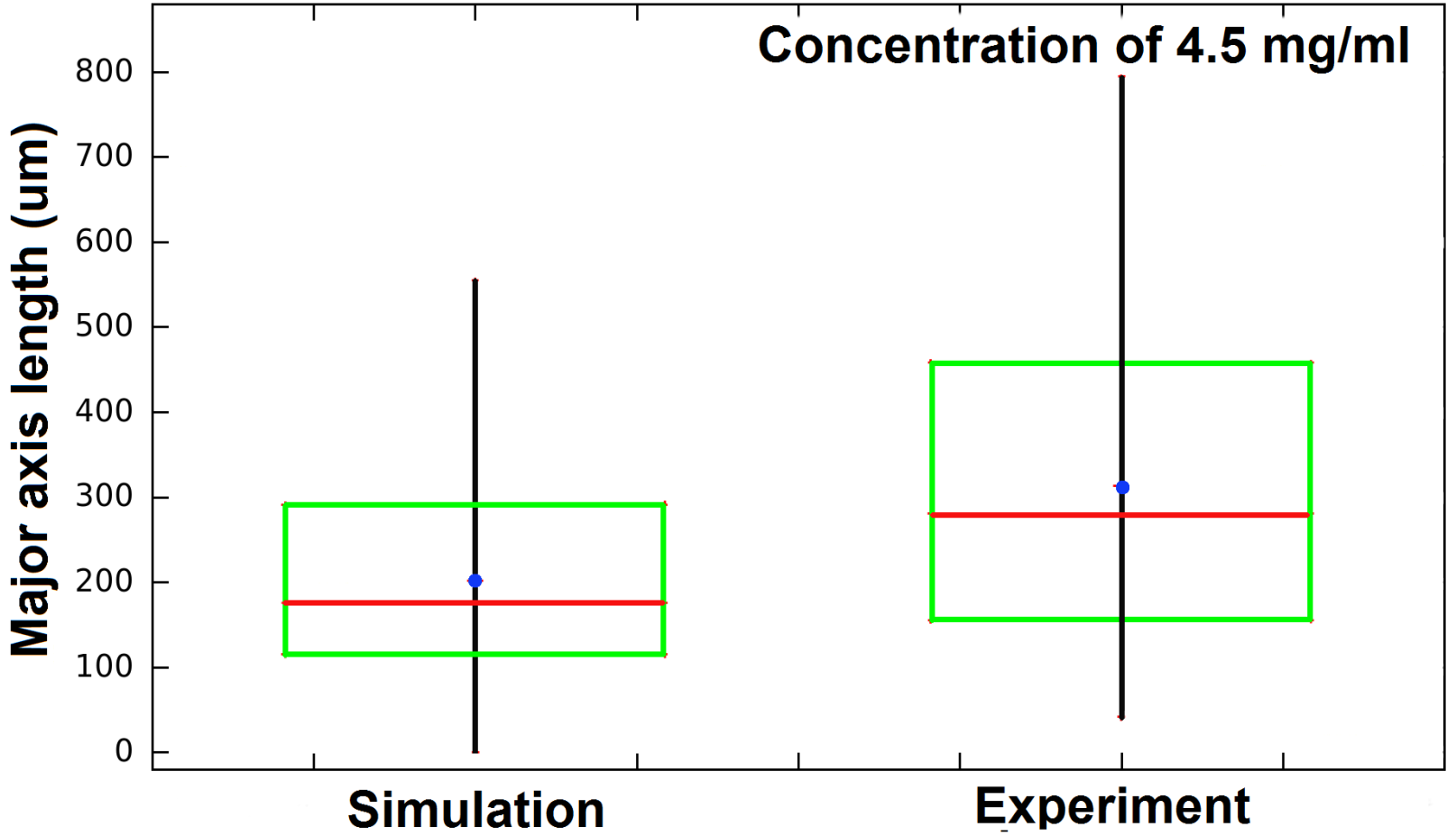}}
\caption{Box plots  show variations in the major axis length of aggregates from the present simulations and the measurements of Ref.~\cite{sixth}. The boxes captures the lower quartile, median (red line) and upper quartile values. The black line that extended from the box show the upper and the lower fence. The blue dots shows the average length of the aggregations in each case. a) $0.563~ mg/ml$, b)  $1.125~ mg/ml$, c) $2.25~ mg/ml$, and d) $4.5~ mg/ml$.}
\label{figure1}
\end{figure}

Due to the external magnetic field, the magnetic particles
are acting like magnets and the particle motion is attributed to gravity and
magnetic interactions forces, since no fluid flow exists. 
The size of the aggregations  that are formed as a function of particle numbers and concentrations is presented in Table \ref{table4}. Results from concentration $1.125~mg/ml,$ for example, show that four aggregates of length $10$ particles may be formed due to the application of this magnetic field. It is observed that the size of the aggregations   is increased as concentration increases and for $4.5~mg/ml$ aggregates of 35 particles may be found. The results from the simulation are compared against the experimental data of Ref.~\cite{sixth} as it is depicted in Fig.~\ref{figure1} and tabulated in Table~\ref{table3} for the mean length of the aggregates and their standard size deviation.

Figure~\ref{figure1} shows the length of aggregates major axis whose average value follows an upward
trend as a function of suspension concentration. The major axis length corresponds
to the longest dimension of the aggregates measured. Larger aggregates are detected in increasing numbers as the concentration of the suspension increases. For the concentration of $0.563~mg/ml$ the numerical results are very close to the experimental, while for the concentration of  $1.125~mg/ml$ results are similar compared to the experimental data in terms of mean length. For the higher concentrations an underestimation may be observed between the results of simulations and the experimental one. This difference is due to the small numbers of particles (up to $380$ particles) that were used in the simulation in comparison in contrast to the measurements that were conducted using a number of particles in the order of $10^6$. 

\begin{table}
\caption{Summary of the results from Case 1}
\label{table4}
\centering
\begin{tabular}{|p{2.9cm}|p{1cm}|p{1cm}|p{1cm}|p{1cm}|p{1cm}|p{1cm}|}
\hline
\multirow{2}{*}{} & \multicolumn{4}{ |c| }{Particles per simulation} \\\cline{2-5}
 & 311 & 385 & 336 & 307  \\
\hline
Particles per aggregation &  0.563 $mg/ml$ &1.125 $mg/ml$ & 2.25 $mg/ml$ & 4.5 $mg/ml$    \\
\hline
1 & 26 & 9 & 0 & 0  \\
\hline
2 & 24 & 13 & 3 & 0 \\
\hline
3 & 18 & 5 & 3 & 1  \\
\hline
4 & 5 & 11 & 3 & 1  \\
\hline
5 & 7 & 5 & 3 & 0 \\
\hline
6 & 7 & 6 & 2 & 0\\
\hline
7 & 7 & 5 & 3 & 0  \\
\hline
8 & 2 & 4 & 0 & 0 \\
\hline
9 & 2 & 2 & 2 & 1  \\
\hline
10 & 1 & 4 & 2 &  1 \\
\hline
11 & 0 & 2 & 11 & 1 \\
\hline
12 & 0 & 2 & 0 & 2 \\
\hline
13 & 0 & 1 & 0 & 0  \\
\hline
14 & 0 & 0 & 1 & 0 \\
\hline
15 & 0 & 0 & 0 & 1  \\
\hline
16 & 0 & 0 & 2 & 0  \\
\hline
17 & 0 & 1 & 0 & 1 \\
\hline
18 & 0 & 0 & 2 & 0\\
\hline
19 & 0 & 1 & 0 & 0  \\
\hline
21 & 0 & 0 & 1 & 0  \\
\hline
23 & 0 & 0 & 1 & 1 \\
\hline
25 & 0 & 0 & 1 & 0  \\
\hline
26 & 0 & 0 & 0 & 2 \\
\hline
27 & 0 & 0 & 0 & 1  \\
\hline
31 & 0 & 0 & 0 & 1  \\
\hline
32 & 0 & 0 & 1 & 1 \\
\hline
35 & 0 & 0 & 0 & 1 \\
\hline
\end{tabular}
\end{table}

\begin{table}
\caption{Comparison of the present results against the experimental measurements from  Ref.~\cite{sixth}.}
\label{table3}
\centering
\begin{tabular}{|c|c|c|c|c|c|}
\hline
Concentr. $(mg/ml)$ &  Particle number & Mean length $(\mu m) $ &  std  & Mean length $(\mu m) $  & std \\
\hline
 - & - & simulation & sim. & experiment & exp. \\
\hline
0.563 & 311 &  34.02 & 23.72 & 31 & 16 \\
\hline
1.125 & 385 &  59.14  & 43.19 & 59 & 36 \\
\hline
2.25 & 336 &  108.93  & 83.79 & 137 & 85 \\
\hline
4.5 & 307 & 201.49  & 109.42 & 317 & 195 \\
\hline
\end{tabular}
\end{table}

The second case that is studied here, which is initially addressed by Vartholomeos and Mavroidis \cite{nineth}, investigates the flow and aggregation of particles under the combine action of a constant and a gradient external magnetic field acting simultaneously. The present results were compared against the experimental measurements and simulations of \cite{nineth} for the motion of a magnetic suspension of polystyrene magnetic particles $(5.5~ \mu m)$ with density of $1050~ kg/m^3$ and  concentration of $25~ kg/m^3$ in distilled water. 

\begin{figure}
\centering


\subfloat[]{\includegraphics[width =6.5cm]{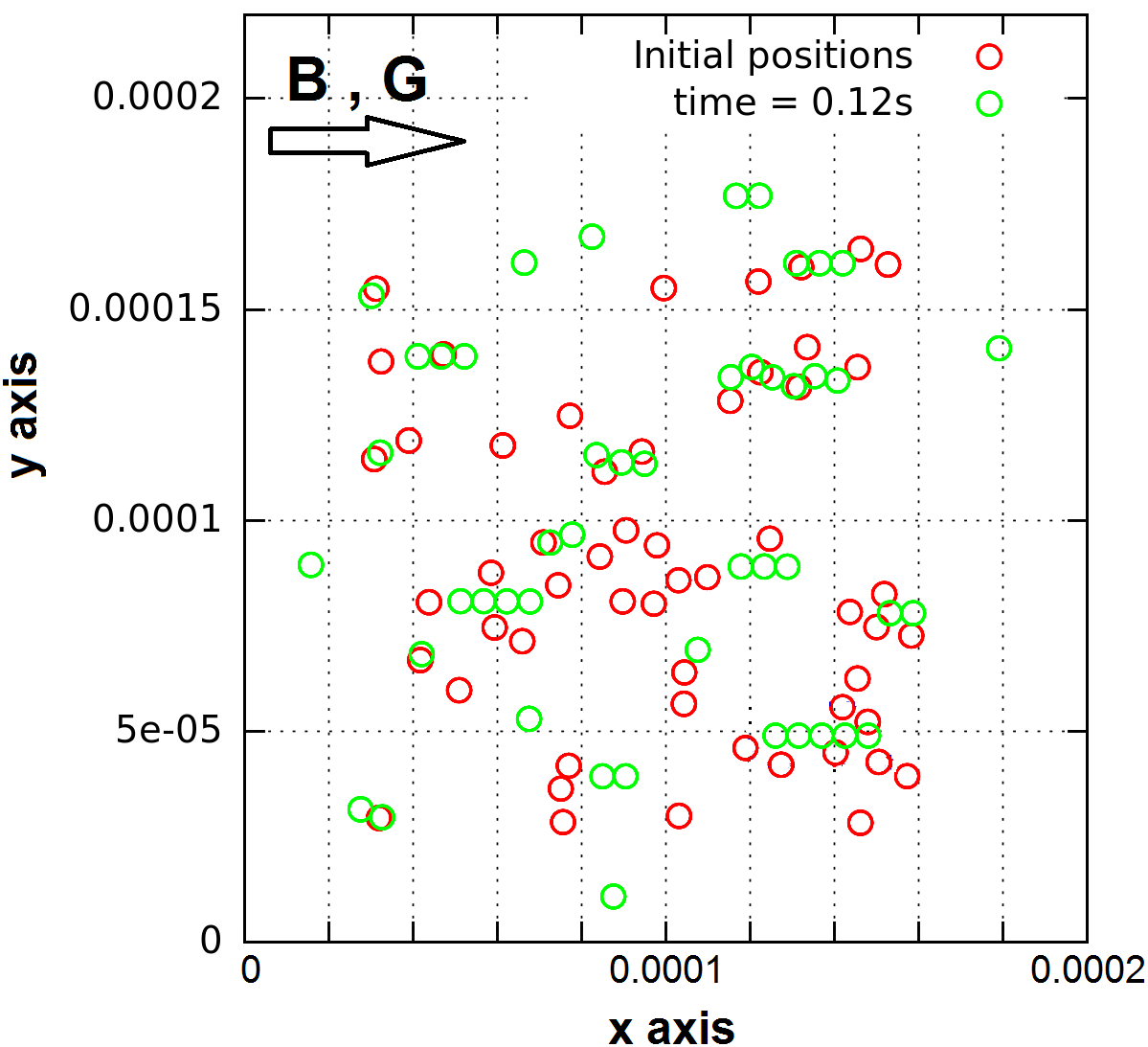}} 
\subfloat[]{\includegraphics[width = 6.5cm]{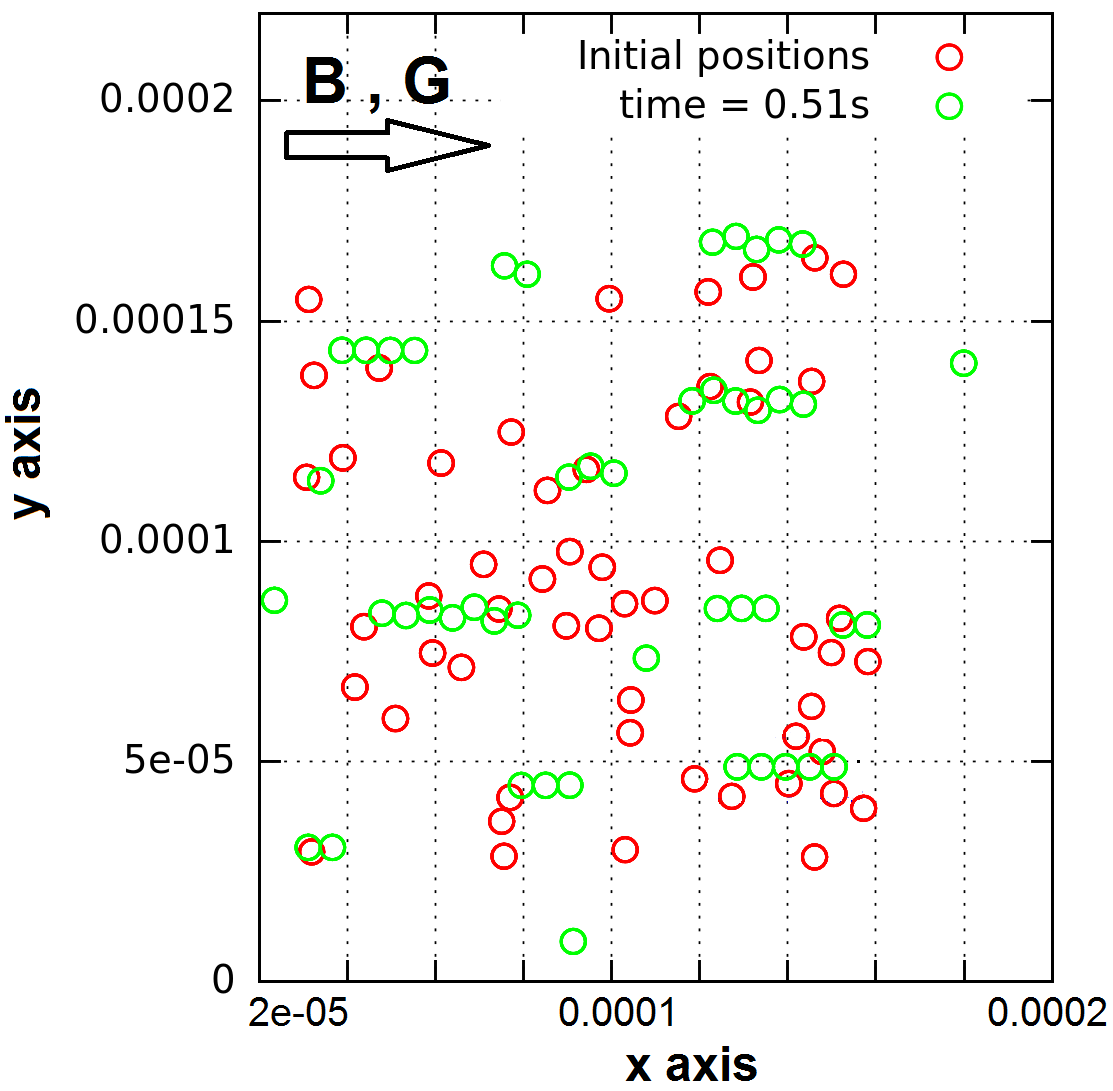}}\\
\subfloat[]{\includegraphics[width = 6.5cm]{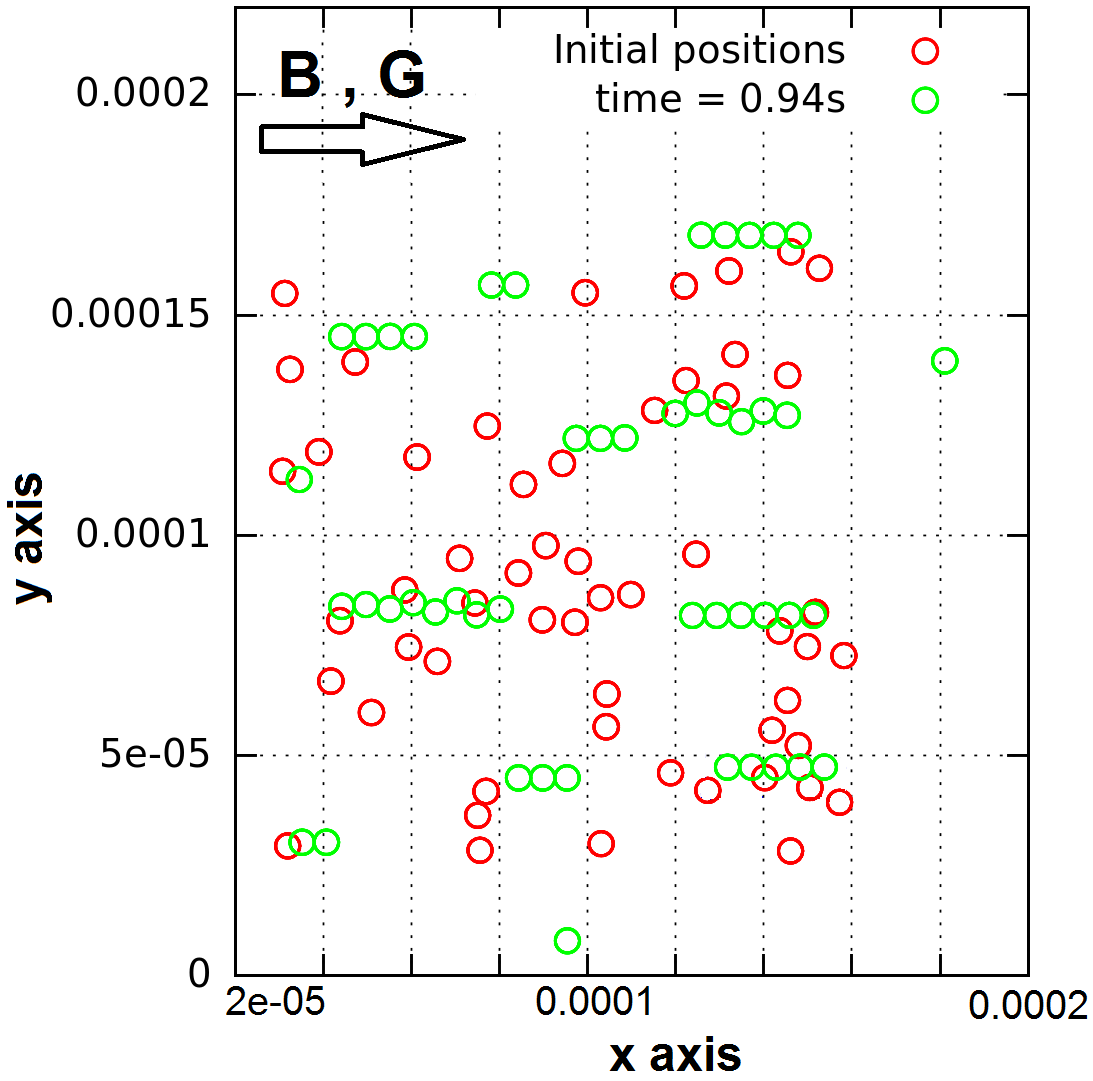}} 
\subfloat[]{\includegraphics[width = 6.5cm]{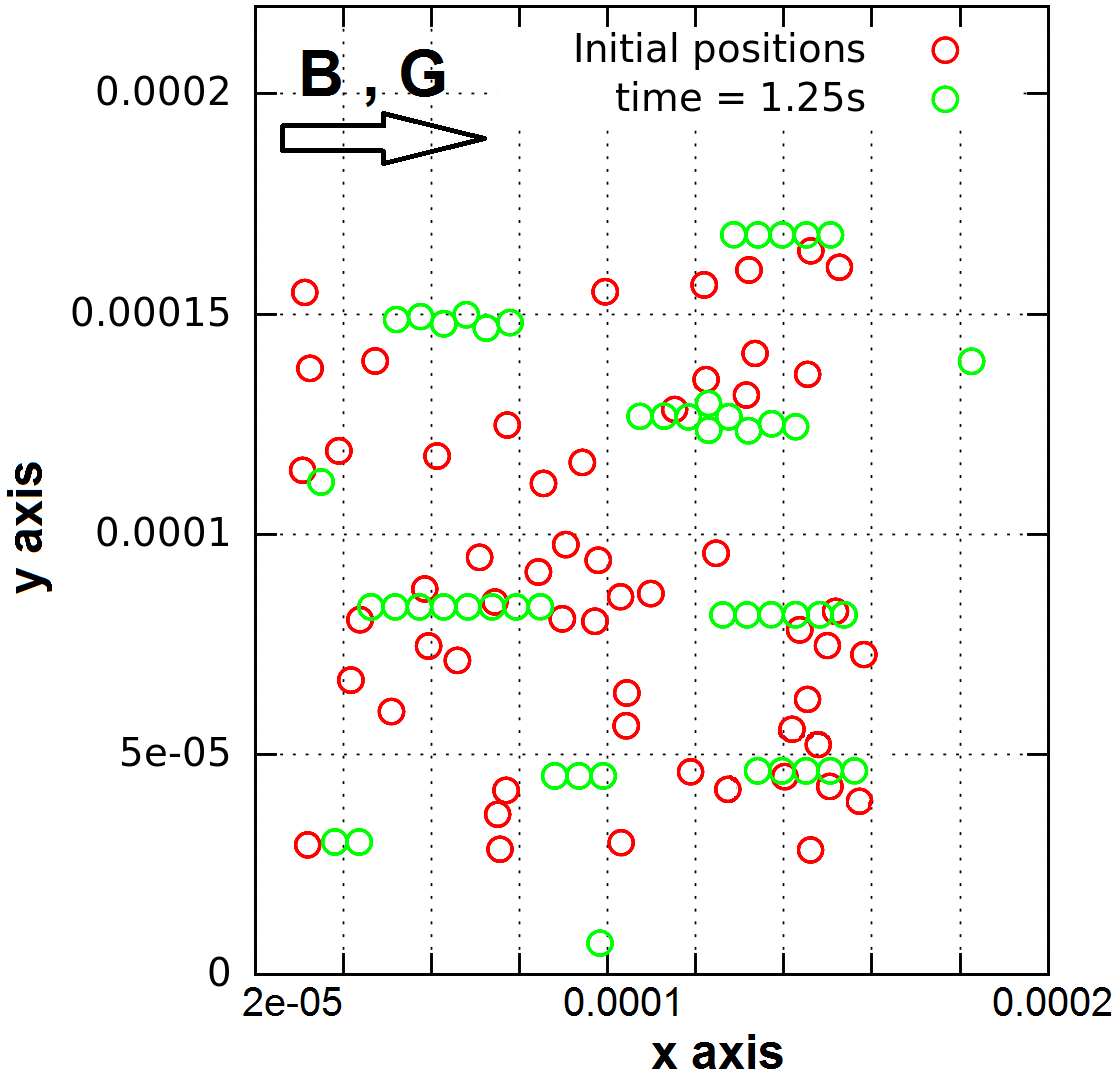}} 
\caption{Staring positions of particles (red circles) and final positions (green circles) for particles displacement between $t= 0~s$ and: a)  $t =0.12~s$,  b)  $t =0.51~s$, c)  $t =0.94~s,$ and d)   $t =1.25~s$.}
\label{fig:figure11}
\end{figure}

Driven by the uniform magnetic field and the magnetic gradient along the $x-$axis, the microparticles are successively aggregated and simultaneously transported in the direction of the magnetic gradient. Four instances of the particle locations that show their motion and the mechanism of the formation of the aggregates are presented in Fig.~\ref{fig:figure11}. Depending of their close neighbors positions, particles may approach others or stay isolated.   The duration for the formation of the aggregates is $t=1.25s$ as depicted in Fig. \ref{fig:figure11}, and for larger times only translational motion of the aggregates is found.
%
Thus, as it is expected, it is found that longer time is needed for the formation of aggregates than in the first case studied due to the slower magnetic forces from the weaker horizontal constant magnetic field. The results of the present simulation  are summarized and compared against the experimental data and simulations of Ref.~\cite{nineth} in Table~\ref{table2}.

It is observed that the present results show good qualitative and quantitative agreement mainly in terms of the mean velocity and the mean length of aggregations in comparison to the experimental one. The small number of particles that is used in the experiment and the simulation
is 
probably the source of the small difference in the comparison because  initial locations of particles may be crucial for the formation and motion of chains.  Moreover, results may be influenced also by the diameter of the particles that was kept constant in the present simulation, while in the experiment had a normal distribution. Consequently, existing small particles were easier and faster concentrated around bigger particles and thus bigger chains of aggregations may be formulated.

\begin{table}
\caption{Comparison of experimental and simulation results for Case 2}
\label{table2}
\centering
\begin{tabular}{|p{3cm}|p{1.4cm}|p{1.3cm}|p{1.8cm}|p{1.5cm}|}
\hline
Case &  Mean size $(particles)$ & std size $(particles)$ & Mean velocity $(\mu m/s)$ & std velocity $(\mu m/s)$ \\
\hline
Experiment, Ref.~\cite{nineth} & 7  & 5 & 7.5 & 1 \\
\hline
Present & 5.26  & 1.89 & 8.3 & 1.4 \\
\hline
Numerical, Ref.~\cite{nineth} & 10 & 4 & 9 & 2 \\
\hline
\end{tabular}
\end{table}

\section{DISCUSSION}

In this study, we were able to validate a new numerical model for the MRI guided drug delivery system that is based on magnetic nanoparticle aggregations for the navigation of paramagnetic nanocupsules. The present method is based on computational fluid dynamics and it is as general as possible and thus can be easily extended for use under realistic in-vitro or in-vivo conditions. The purpose   of this work is to examine under which conditions the model is reliably working. For this reason, two basic particle suspension flows were selected where particles are influenced by a constant and in the second case a gradient magnetic field. 

Simulations with the present method is more expensive as the number of particles increases because magnetic moments should be calculated for every particle and for every time step. However, given a particle suspension and the magnitude of the external magnetic field the method can be applied and provide very good accuracy for the prediction of the size of conglomerates. As it is observed from Table~\ref{table3}, when the ratio of the number of the particles to the suspension concentration is high, numerical results are very close to the experimental. As this ratio decreases,  the method underestimates the size of the aggregations but still is capable to predict qualitatively their growth.  

An important advantage of this work is also that the developed model is found to predict more accurate the dynamics of the aggregates than other simulations with similar models as Table~\ref{table2} presents. Results show that the most important quantity for the driving of the nanoparticles, i.e. the mean velocity which is developed under gradient magnetic fields, is very closely predicted to the experimental values. This improved feature  of the proposed model is due to the evaluation of the more realistic aerodynamic coefficient $C_dA_{eff}$ that incorporates the effect in the drug from the exposed shape of the aggregates in a better way, since it is general true that $C_dA_{eff}<C_dA$.

\section{CONCLUSION}

A numerical model for the simulation of magnetically guided drug delivery was developed in order to predict the motion of magnetic particles for medical applications. This method aims at predicting the aggregation's procedure and the velocity of each particle inside the arteries and arterioles of a human body. It aims also at simulating the motion of aggregations when these are driven by MRI magnetic gradient coils.
A novel method is developed for the calculation of the drag coefficient that takes into account the exposed area of the particle in the fluid.

The method was tested through comparison against experimental and numerical data. It was found that the present method can simulate satisfactory the experiments in a stationary fluid under a steady magnetic field. Furthermore, the model was tested for the acceleration of aggregated particles  under the influence of a constant and a superimposed gradient magnetic field. The results were very close to existing experimental data in terms of velocity and aggregation size and comparable to the results from existing simulations.

\section*{Acknowledgments}
\addcontentsline{toc}{section}{Acknowledgment}

The work is funded by the NANOTHER program (Magnetic Nanoparticles for targeted MRI Therapy) through the Operational Program COOPERATION 2011 of GSRT, Greece.
Discussions with Dr Klinakis from BRFAA, Greece, Prof. Zergioti from NTUA, Greece and the people from Future Intelligence Ltd. are also acknowledged. 


\end{document}